\documentclass[runningheads,a4paper]{llncs}

\usepackage{amssymb}

 \usepackage{url}
 \urldef{\mailsa}\path|{dfay.bournemouth.ac.uk}|
 
 \newcommand{\keywords}[1]{\par\addvspace\baselineskip\noindent\keywordname\enspace\ignorespaces#1}

 \usepackage{amsmath}
 \usepackage{multirow}
 \usepackage{graphicx}
 \usepackage{graphics}
 \usepackage{color}
 \usepackage{tabularx}
  \usepackage{cite}

  \DeclareMathOperator*{\argmin}{arg\,min}
  
 \usepackage{floatrow}
 \usepackage{blindtext}
 \usepackage{wrapfig}
  \usepackage{subfig}
 \usepackage{afterpage}
 
 \usepackage{caption}
 \DeclareCaptionType{copyrightbox}

\begin{document}
\mainmatter
\title{Predictive partitioning for efficient BFS traversal in social networks.}

\titlerunning{Predictive partitioning for efficient BFS traversal in social networks.}

\author{Damien Fay}

\authorrunning{D. Fay}

\institute{Department of Computing, Bournemouth University,\\
United Kingdom\\
\mailsa\\}

\toctitle{Lecture Notes in Computer Science}
\tocauthor{Authors' Instructions}
\maketitle


%



\begin{abstract}
In this paper we show how graph structure can be used to drastically reduce the computational bottleneck of the Breadth First Search algorithm (the foundation of many graph traversal techniques). In particular, we address parallel implementations where the bottleneck is the number of messages between processors emitted at the peak iteration. First, we derive an expression for the expected degree distribution of vertices in the frontier of the algorithm which is shown to be highly skewed. Subsequently, we derive an expression for the expected message along an edge in a particular iteration. This skew suggests a weighted, iteration based, partition  would be advantageous. Employing the METIS algorithm we then show empirically that such partitions can reduce the message overhead by up to 50\% in some particular instances and in the order of 20\% on average. These results have implications for graph processing in multiprocessor and distributed computing environments.   


\end{abstract}





%
%

\section{Introduction}
\label{sec:intro}

Breadth First Search (BFS) is a fundamental graph algorithm which is applied constantly to huge social network graphs in distributed and parallel systems consuming large amounts of energy and resources. BFS is central to several more complicated graph algorithms such as identifying connected components, testing for bipartiteness, belief propagation, finding community structures in social networks and computing the max flow-min cut for a graph\cite{Merrill_2012}. As such it has drawn much attention from from the parallel processing community as a benchmark algorithm with several competing variants focused on efficient implementation \cite{Merrill_2012,bulucc2013graph,Kowalski,Shang13,Chen_powerlyra,Z_FU,YangzihaoWang,BaderGraph,Luo_Lijuan,Bisson,Aydin,Yuan_Liang,Buluc_Aydin}. However, despite its importance \emph{known structural properties} of social networks have not been leveraged to improve the algorithms efficiency. We show that a simple adjustment of the partitioning vector based on graph structure can radically improve the efficiency at the algorithms bottleneck, and have little (sometimes improved) effects elsewhere. 

The setting here envisages that BFS is performed repeatedly on an unweighted, undirected graph from random root vertices. In addition, we assume basic statistics about the graph can be collected after each run or alternatively offline. It is assumed the graph is traversed in parallel by several processors thus requiring a-priori a partition of the graph vertices across each processor. In this setting the basic computation step of BFS (Section~\ref{subsec:BFS}) is dominated by the communications costs (messages) between processors after each iteration (as noted in ~\cite{Z_FU} amongst others). The messages emitted after the peak iteration further dominate the communication costs amounting to $\sim70\%$ of the total (Section~\ref{sec:results}), thus this is the \emph{bottleneck} of the whole algorithm. 

The aim of this paper is to use known properties of social networks (specifically the power-law distribution, small world and assortativity properties) to reduce the communications costs at the bottleneck and so make the algorithm more efficient. With the exception of a few papers (Section~\ref{sec:related}) most approaches ignore information about the structure of the graph focusing instead on CPU-GPU architecture specifics. We show that the incident degree distribution per iteration is highly skewed away from a power law distribution. Thus the number of edges crossing a partition is a biased estimate of the messages between partitions at the peak iteration. Further we propose a new weighted graph construction which reflects the expected number of messages per edge. Finally, we show empirically that using the METIS~\cite{metis} partitioning algorithm that the subsequent reduction in messages emitted across partitions can be reduced in some individual cases by $\sim50\%$, for some graphs on average 
by $\sim20\%$ but the improvement is highly dependent on structure and so for some graphs is insignificant. 

The paper is laid out as follows. Section~\ref{sec:related} discusses related work, Section~\ref{sec:background} gives the background behind the BFS algorithm, partitioning and develops the theory showing that the degree distributions are highly skewed. Section~\ref{sec:results} presents empirical results and finally in Section~\ref{sec:conclusion} we mainly focus on future work and discussing the consequences of the findings.

\section{Related Work}
\label{sec:related}

Implementing BFS in parallel is a well established approach which generally consists of three stages: graph pre-ordering, graph partitioning and parallel architecture specific implementation. This research is most pertinent to graph partitioning however there are several aspects of architecture specifics of interest. 

Graph partitioning seeks to reduce the number of messages sent between partitions during processing which can be achieved in several ways. The most obvious mechanism is to use a 1-D partition; each vertex and associated edges are sent to an individual processor~\cite{Merrill_2012,Idwan,Luo_Lijuan,Shang13}. An excellent overview of 1-D graph partitioning methods can be found in~\cite{Buluc_Aydin} with techniques designed specifically for scale-free networks exist such as~\cite{Abou-Rjeili}. Although~\cite{Abou-Rjeili} considers partitioning for social network graphs they do not do so in the context of BFS, indeed the two approaches are complimentary as here we provide a weighted social network graph for partitioning.

Shang and Kitsuregawa~\cite{Shang13} consider partitioning edges across processors (as opposed to vertices). The edges may be uniformly distributed by either the source or the target vertex. They propose that when the degree of the target vertex exceeds a pre-defined threshold the algorithm performs best by switching to a target vertex partitioning, while Hong et. al.~\cite{Sungpack_HONG} note that for low degree vertices partitioning should be based on vertex but for large degree vertices the partitioning should be based on edge. In contrast a 2-D partition~\cite{BaderGraph,bulucc2013graph,Bisson,Aydin} distributes the edges of a vertex across several processors. The 2-D approach is based on the observation that an exploration from a set of vertices is equivalent to the product of the adjacency matrix and a vector of the vertices touched. Thus they partition the adjacency matrix into two dimensions (blocks along the rows and columns) and then collect the row products in one set of messages and the unique column entries in another. Thus the messages produced are between particular processors and not \emph{all to all} as in the 1-D case. It would appear from the literature that the 2D partitioning approach results in more efficient BFS traversals but we do not consider this approach in this research (see future work, Section~\ref{sec:conclusion}).

Skewed graph structure is a central topic in many papers~\cite{Chen_powerlyra,Merrill_2012,bulucc2013graph,Sungpack_HONG,Yuan_Liang}. The non-locality of neighbours in a graph, and the fact that some vertices can have degrees several factors larger than the average, leads to load imbalances across processors and random memory access patterns. Yuan et. al.~\cite{Yuan_Liang} examines the expected distance between two pairs of nodes being explored in a BFS and show that they can predict the vertex locality. This is perhaps the closest work to this research. In contrast our approach looks at the expected use of a vertex of a given degree in a particular iteration, though the two approaches are similar in spirit. To the best of our knowledge our approach is the first to take graph structure during the execution of the BFS algorithm into account.  

\section{Background}
\label{sec:background}

\subsection{Breadth First Search}
\label{subsec:BFS}

\begin{figure}[h]
\centering
\includegraphics[scale=0.4]{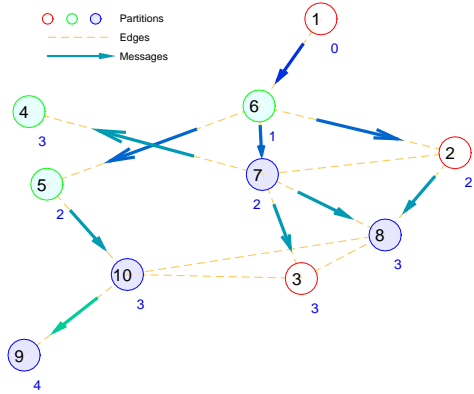}
\caption{Illustration of BFS on a small graph with (arbitrary) partitions. }\label{f:small_graph}
\end{figure}

Given a graph $G(V,E)$ and a source vertex $s$, where $V$, $E$ refer to the vertex and edge sets respectively the BFS algorithm returns a route from $s$ to every reachable vertex in $G$. The BFS algorithm begins with a set $V_0 = \{s\}$ and explores the graph by identifying neighbours of $s$, denoted as the set $V_0^{+}$, where + denotes neighbour expansion. At the next iteration all vertices connected to $V_0^{+}$ minus those already visited are $V_1 = V_0^{+}\setminus \{V_0\} $. We call the set of unique vertices in the $\tau^{th}$ iteration, $V_\tau$, the \emph{frontier} set. In general the frontier consists of 
\begin{equation}
V_{\tau} = V_{\tau-1}^{+}\setminus \{\bigcup_{i=0}^{\tau-1} V_i\}  
\end{equation}
and the set of vertices already visited, $\{\bigcup_{i=0}^{\tau-1} V_i\}$, are said to be \emph{touched}. The algorithm continues until $V_{\tau} =\{ \emptyset \}$ and all vertices have been explored. Figure~\ref{f:small_graph} illustrates a small example starting vertex 1, where $V_0 = \{1\}$, $V_1 = \{6\}$, $V_2 = \{2,7,5\}$, $V_3 = \{4,10,3,8,8\}$, $V_4=\{9\}$ and finally $V_5 = \{ \emptyset \}$. In this example we see that the number of vertices in the frontier increases rapidly, peaks, and then decays rapidly to zero. In addition, note that there is a duplicate in $V_3$ as vertex 8 is reached from both vertex 7 and 2 in the $2^{nd}$ iteration. The BFS algorithm creates a shortest path tree from the root node by recording the edges traversed between $V_{\tau}$ and $V_{\tau-1}$ and in the case of duplicates, only the first (or a random) edge is recorded. 

The BFS algorithm may be implemented on $P$ parallel processors by partitioning $V$ into $P$ subsets $\mathcal{V}_1, \ldots, \mathcal{V}_{P}$, where $\mathcal{V}_i \bigcap \mathcal{V}_j = \{\emptyset\} \quad \forall i \ne j$, and $ \bigcup_{i} \mathcal{V}_{i} = V $, such that each vertex is assigned a processor which performs the neighbour expansion of that vertex. This is the basic format of most parallel BFS (P-BFS) algorithm implementations. At the end of each iteration the processor owning each element in the next frontier must be notified that this vertex is now to be explored. We define a message $\mathcal{M}_{\mathcal{V}_i \rightarrow \mathcal{V}_j}^{\tau}(u,v)$ to be a notification from processor $i$ to processor $j$ that vertex $u$ has identified vertex $v$ to be a member of the next frontier set. If $u$ and $v$ reside in the same processor then there is no communication cost and thus the communications cost for P-BFS is here defined as the sum of all messages that cross a partition: 
\begin{equation}
 C^{\tau} =  \sum_{u\in V_{\tau-1},v \in V_{\tau}}  \mathcal{M}_{\mathcal{V}_i \rightarrow \mathcal{V}_j}^{\tau}(u,v) \quad  \forall i \ne j
\end{equation}

This is illustrated in Figure~\ref{f:small_graph} where the graph is arbitrarily partitioned into 3: $\mathcal{V}_1 =\{1,2,3\}, \mathcal{V}_2 =\{4,5,6\},\mathcal{V}_3 =\{7,8,9,10\}$. In this case the messages in iteration 2 of the algorithm are $\mathcal{M}_{\mathcal{V}_2 \rightarrow \mathcal{V}_3}^{2}(7,4)$, $\mathcal{M}_{\mathcal{V}_3 \rightarrow \mathcal{V}_1}^{2}(7,3)$, $\mathcal{M}_{\mathcal{V}_1 \rightarrow \mathcal{V}_3}^{2}(2,8)$, $\mathcal{M}_{\mathcal{V}_2 \rightarrow \mathcal{V}_3}^{2}(5,3)$. 

The aim of a partitioning algorithm is to partition a graph into roughly equal sets, $|\mathcal{V}_i| \approx |\mathcal{V}_j|$, such that a specific objective is achieved such as the number of edges that cross the partitions, the \emph{edge-cut}, is minimized as: 
\begin{equation}
 \argmin_{\mathcal{V}_1,\ldots,\mathcal{V}_P} C =  \sum_{u\in \mathcal{V}_i,v \in \mathcal{V}_j, \forall i \ne j}  w_{u,v} 
\end{equation}
or alternatively the communications volume: 
\begin{equation}
\argmin_{\mathcal{V}_1,\ldots,\mathcal{V}_P} C =  \sum_{u\in \mathcal{V}_i \forall i }  w_{u,v} D(v)
\end{equation}
where $D(v)$ is the number of different blocks in which $v$ has a neighbouring node (to account for parallel communications between processors). Although the communications volume would appear to be more suited as a objective function for P-BFS the minimum cut has been adopted as the standard for several practical reasons~\cite{Buluc_Aydin}. There are several methods for graph partitioning (a recent review of such methods may be found in~\cite{Buluc_Aydin}) and the one adopted here is the popular METIS~\cite{metis} multi-level k-way algorithm. This algorithm first applies several rounds of coarsening to produce a small graph which is then partitioned. This partitioned graph is then uncoarsened in stages to produce the full partition. 

\subsection{Graph structure}
\label{subsec:graph_structure}

The development here initially follows that of Kurant et. al.~\cite{Kurant} who derive expressions for the \emph{observed} degree distribution of a graph sampled by BFS (i.e. a different problem). The configuration model~\cite{orbis} is a construct which allows construction of graphs with a desired degree distribution. $N$ vertices are each assigned $k$ \emph{stubs} sampled uniformly from a desired degree distribution, $p_k$, i.e. $k\sim p_k$. The configuration model then pairs these stubs at random thus constructing edges and thus a graph with the desired degree distribution. The order in which these stubs are connected is irrelevant as the pairing is random. Thus we may assign to each stub an arbitrary time, $t \in [0,1]$ and moving from $t=0\rightarrow 1$ connect the stubs as their randomly assigned time is passed. This converts a discrete graph generation process into a continuous time process and is a useful framework to derive expressions for the bias inherent in BFS sampling~\cite{Kurant,Achlioptas}. 
Kurant et. al. interweave the stub matching step with the exploration phase of BFS. Thus the stubs are connected only when the the frontier is being explored and the unconnected stubs with the lowest time are chosen first. A vertex enters the frontier when all of its stubs have been paired and this happens with probability ($1-t)^k$ therefore the expected fraction of vertices of degree $k$ touched before time $t$ is~\cite{Kurant}:
\begin{equation}\label{e:f_k}
 f_k(t) = p_k(1-(1-t)^k)
\end{equation}
where $p_k$ is the probability a vertex has degree $k$ (i.e. the degree distribution). The fraction of nodes of any degree visited before time $t$ is~\cite{Kurant}:
\begin{equation}\label{e:fract_total}
 f(t) = 1-\sum_k p_k(1-(1-t)^k)
\end{equation}
Kurant investigated the bias of BFS samples but here we are concerned with the degree distribution of the frontier. In addition, we are only interested in particular times; those that correspond to the iterations. It is assumed that the number of vertices touched up to each iteration $n_{\tau} = \sum_{i=1}^{\tau} |V_{i}|$ is known.\footnote{A good estimate of the number of vertices expected in each iteration of BFS can be obtained from a single graph traversals.} Thus we may define:\footnote{Here we diverge from the analysis of Kurant as we are interested in the degree distribution used per iteration and not the degree distribution of the BFS tree.} 
\begin{equation}
 t_{\tau} = f^{-1}(n_{\tau}/N) 
\end{equation}
where $f^{-1}$ denotes the inverse of $f(t)$ as (\ref{e:fract_total}) cannot be inverted explicitly. This inverse consists of finding the minimum of a smooth function in one dimension and may be solved easily using gradient descent or any similar search algorithm. The frequency of degrees of type $k$ in the $\tau^{th}$ frontier, $n_k^{\tau}$, can be calculated iteratively by removing those seen in the previous frontiers as: 
\begin{equation}
 n_k^{\tau} = \frac{p_k(1-(1-t_{\tau})^k)}{\sum_k p_k(1-(1-t_{\tau})^k)} n_{\tau} - \sum_{i=1}^{\tau-1}p_k^{i} n_{i} 
\end{equation}
where $n_{0}=0$ and $p_k^{\tau}$ is the frontier degree distribution defined as: 
\begin{equation}
 p_k^{\tau} = \frac{n_k^{\tau}}{\sum_k n_k^{\tau}}
\end{equation}
The probability that a vertex of degree $k$ is used in the frontier is then the number of vertices of degree $k$ in the frontier divided by the total number in the graph: 
\begin{equation}\label{e:pi_k}
 \pi_k^{\tau} = \frac{p_k^{\tau}n_{\tau}}{p_kN}
\end{equation}

\begin{figure}
\begin{floatrow}
\ffigbox{  \includegraphics[width=0.45\textwidth]{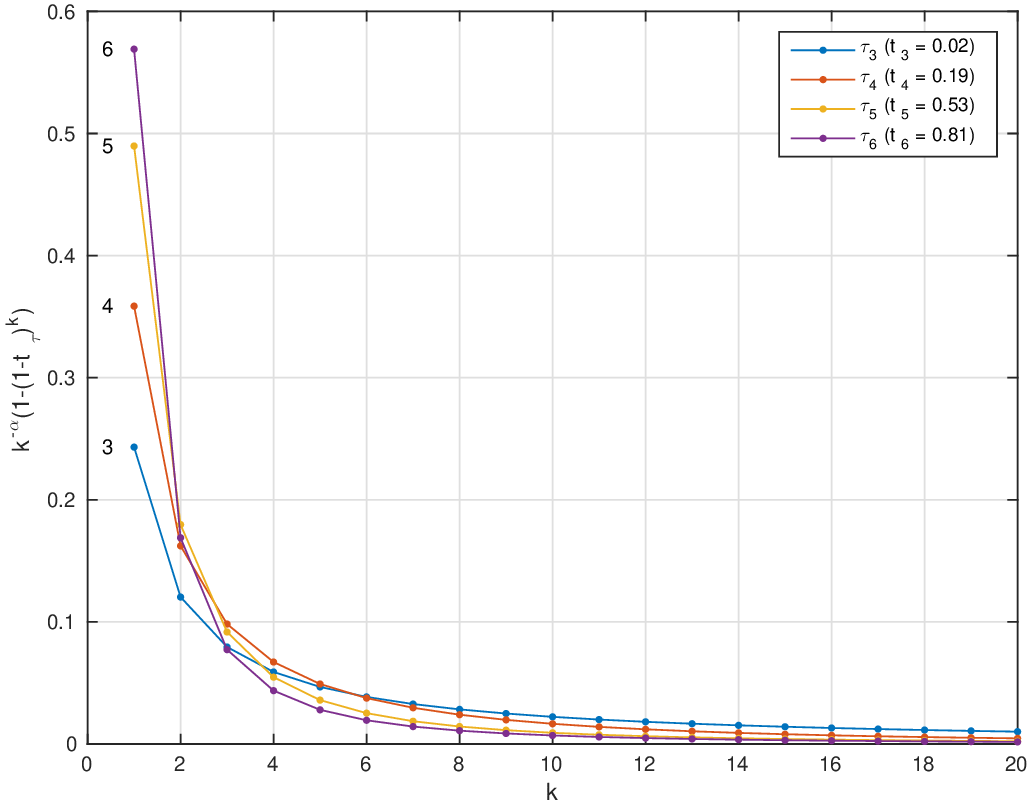} }
{%
\caption{Proportion of vertices of degree $k$ seen before iteration $\tau$ ($\alpha = -2$)}\label{f:fk_tau_theoretical}
}
\ffigbox{  \includegraphics[width=0.45\textwidth]{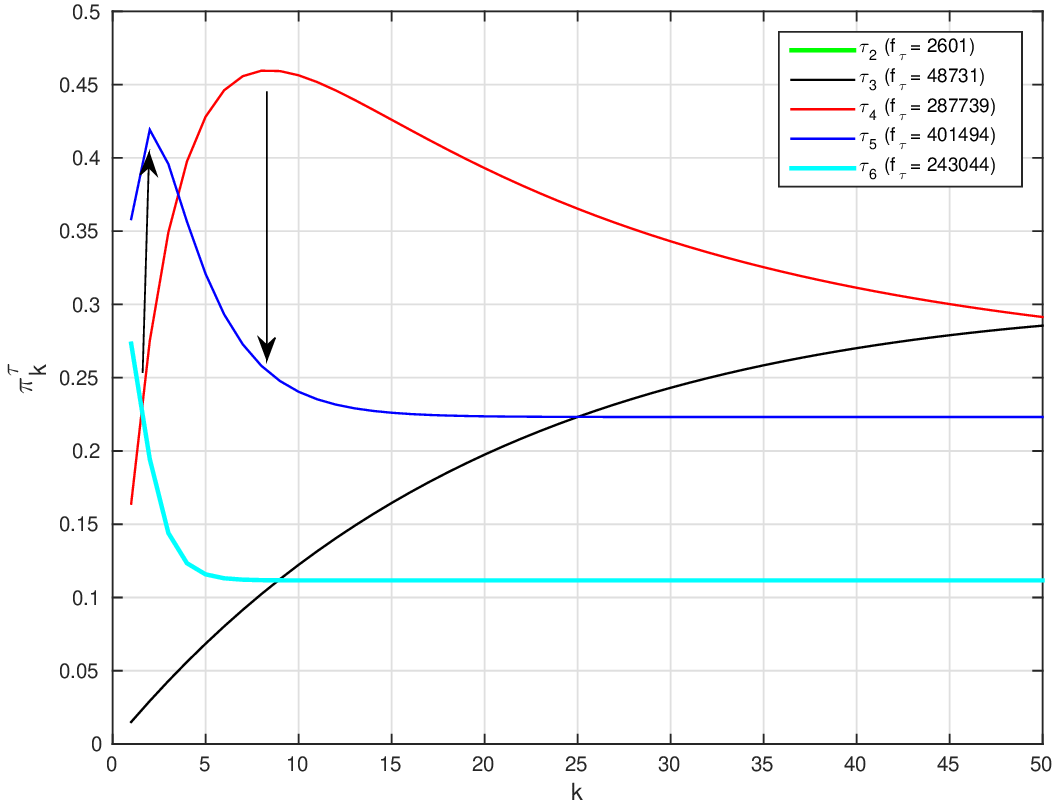} }
{%
\caption{Probability a vertex of degree $k$ will be used in iteration $\tau$ (theoretical)}\label{f:pi_tau_theoretical}
  
}

\end{floatrow}
\end{figure}

Figure~\ref{f:fk_tau_theoretical} shows $f_k(t)$ for $p_k \propto k^{-2}$.\footnote{Here we use the YouTube friendship graph as an example: the power law exponent =-2 and $t_\tau$ = $\{0.0006,  0.02,0.19,0.53,0.81,0.93,0.97,0.99,1\}$, the results are similar for the other graphs we examined.} Up to iteration 3, $25\%$ of the degrees touched are of degree 1 which rises to $\sim 50\%$ by iteration 5. That is, BFS is biased (proportionately) towards higher degree vertices initially, moving towards lower degree vertices at later iterations. Note that Figure~\ref{f:fk_tau_theoretical} shows the \emph{accumulated} proportion as the algorithm progresses, however, it is the difference in these proportions that are touched at each iteration and this has a very different shape (Figure~\ref{f:pi_tau_theoretical}). 

Figure~\ref{f:pi_tau_theoretical} shows $\pi_k^{\tau}$ for the YouTube friendship graph (Section~\ref{sec:results}). The distribution of nodes used in iterations 2 and 3 is biased towards high degree nodes. In iteration 4 the bias centres on vertices of degree 10 with $40\%$ being touched but only $15\%$ of degree 1 nodes are touched. In iteration 5 the bias switches, $\sim40\%$ of degree 1 vertices are touched but only $\sim25\%$ of degree 10 vertices are touched. There is a similar switch between iteration 5 and 6. The interesting thing about this behaviour is that the degree distribution of vertices used is highly skewed and during the main iterations (4,5,6) those used in one iteration tend not to be used in the next and visa versa (as illustrated with arrows in Figure~\ref{f:pi_tau_theoretical}). Thus at a specific iteration we have a prior probability over the vertices that will be used \emph{and} a different prior over the vertices they are connected to in the next frontier, and these distributions 
are different from the initial power-law distribution, i.e. $ \pi_k^{
\tau} \ne \pi_k^{\tau+1} \not\propto p_k$. 
The transition from $\pi_k^{\tau} \rightarrow \pi_k^{\tau+1}$ involves connecting vertices with degree distribution $\pi_k^{\tau}$ to those with $\pi_k^{\tau+1}$. It would be tempting to assume that the probability of a node of degree $k$ connects to a node of degree $k'$ is just the product of $\pi_k^{\tau}$ and $\pi_k^{\tau+1}$, however the two events are not independent. Real-world graphs are generally assortative and as has been shown graph generators that take into account the correlation structure in the joint degree distribution $p_{k,k'}$ produce far better approximations to real-world graphs~\cite{orbis} and have very different properties from those that assume independence~\cite{Fay20113458}. Here we assume that the joint degree distribution, $p_{k,k'}$,~\cite{orbis} gives a good approximation of the expected edges between the vertices in iteration $\tau$ and $\tau+1$, therefore we may define the probability of transitioning from a vertex with degree $k$ to an edge with degree $k'$ in iteration $\tau$, $p_{k,k'}^{\tau}$ as: 
\begin{equation}\label{e:pkk}
  p_{k,k'}^{\tau}  = \pi_k^{\tau} p_{k,k'} \pi_k'^{\tau+1} 
\end{equation}
The probability of using a particular edge, $\{u,v\}$, in iteration $\tau$ is equal to the probability of passing from $u \rightarrow v$, or from $v\rightarrow u $ but not both, $u\leftrightarrow v$, as this would imply $u$ and $v$ have already been touched in iteration $\tau$, therefore:
\begin{equation}\label{e:wkk}
  w_{k,k'}^{\tau}  = p_{k,k'}^{\tau} + p_{k',k}^{\tau} - p_{k,k'}^{\tau}p_{k',k}^{\tau}
\end{equation}
where $w_{k,k'}^{\tau}$ can be used to weight each edge in $G$ where the weights represents the expected message along that edge in iteration $\tau$. The total number of expected messages given a particular partition is then: 
\begin{equation}\label{e:weighted_cut}
 E[C^{\tau}] =  \sum_{u\in V_{\tau},v \in V_{\tau+1}}  w_{k_u,k_v}^{\tau}  \mathcal{I}_{\mathcal{V}_i \rightarrow \mathcal{V}_j}(u,v) \quad
\end{equation}
where $\mathcal{I}_{\mathcal{V}_i \rightarrow \mathcal{V}_j}(u,v)$ is an indicator variable s.t. $u\rightarrow v$ crosses a partition. To implement this approach requires estimates of; $p_k$, $p_{k,k'}$, $n_\tau$. Given these a weighted version of $G$, $W(V,E)$, may be constructed, and partitioned using a weighted partitioning algorithm (here we use METIS). 






\section{Results}
\label{sec:results}
The simulations presented below consist of randomly choosing a source node, performing a BFS using the competing algorithms (described below), and recording the number of messages generated. The simulations are based on 500 randomly chosen root vertices. We begin by looking at the messages generated in individual runs, moving onto those for a particular graph and finish with the results of the simulations over the 8 graphs examined. The \emph{peak iteration} is defined as the iteration with the largest number of vertices in the frontier, i.e.. $max_{\tau} |V_\tau|$,  There are four competing algorithms which represent different levels of knowledge: 

\begin{enumerate}
\item METIS is applied to G(V,E) from $s$ with no weighting, a BFS is then performed and the messages across partitions recorded. This is the \emph{baseline} algorithm against which the others are compared,
\item Using the results from 1) we calculate the peak $\tau$, and $p_{k,k'}^{\tau}$ using actual messages counts. $G$ is then weighted to give the empirical weighted matrix which we denote:  $W_{emp}$. $W_{emp}$ is then partitioned using METIS. This partition is then used to perform the same BFS from $s$. Note that in essence we are using the answer to derive the partition which is unrealistic. The aim here is to give an upper bound on the algorithms performance, 
\item Using the $p_{k,k'}^{\tau}$ from all 500 iterations we combine and smooth these estimates to produce a single weighted graph called, $W_{smooth}$. This is partitioned using METIS and a BFS is performed from $s$. The aim here is to give an estimate of performance without the approximation error inherent in Equation~\ref{e:pi_k}, and 
\item Using the actual degree distribution, $p_k$, the joint degree distribution $p_{k,k'}$ (see note below) and the number of vertices in the peak iteration together with equations (\ref{e:pi_k},11,12) we form a single weighted graph called, $W_{avg}$. We note that these quantities are computationally inexpensive to calculate and a reasonable estimate may be formed from a small number of BFS runs (here we use 10 runs). 
\end{enumerate}

The joint degree distribution, $p_{k,k'}$, can present problems of storage and estimation especially when the maximum degree is high. However, as the graphs studied have a power law distribution, the number of vertices with a high degree falls rapidly. In this paper we calculate $p_{k,k'}$ where nodes with $k \ge 300$ are counted in a single bin. Therefore, $p_{k,k'}$ is formed of a, $300\times300$ grid. We choose the number of partitions to be be $100$ as this reflects the order of processors in a GPU (the number of processors varies greatly depending on the machine; the NVIDIA GeForce GTX280, for example, has 30 \cite{Luo_Lijuan} while the NVidia Kepler architecture has 4,096 GPU's\cite{Bisson}). 

\subsection{Data Sets}
\label{subsec:datasets}
The datasets used in this study are taken from the Konect graph repository.\footnote{http://konect.uni-koblenz.de} We are specifically interested in social network graphs and so the RMAT graphs used in studies such as~\cite{Merrill_2012,YangzihaoWang} are not included though we do include a synthetically generated ER graph with a single large component. We also did not consider graphs with $N> 2$M for computational reasons. These graphs are listed in Table~\ref{t:congruence}.

\subsection{Simulations}
\label{subsec:SpecificCase}
\begin{figure}
\begin{floatrow}
\ffigbox{  \includegraphics[width=0.45\textwidth]{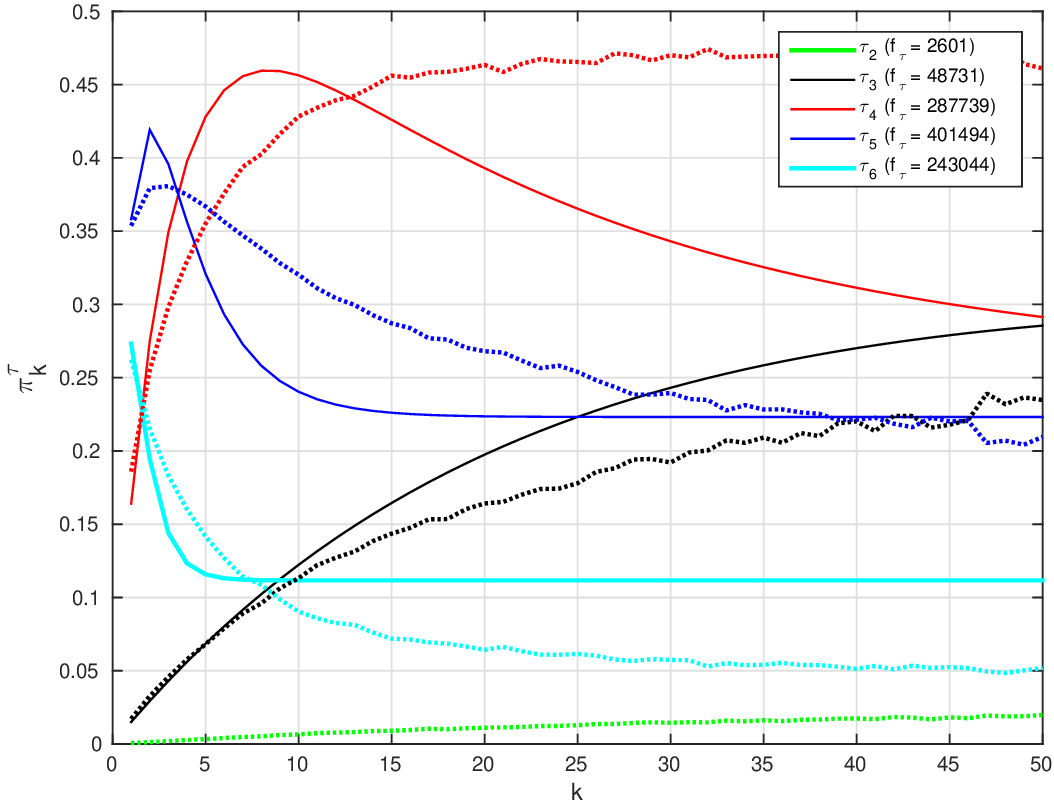} }
{%
\caption{$\pi_k^{\tau}$ theoretical (solid) vs empirical (dashed) (YouTube Graph).}\label{f:pi_tau_theoretical_actual}
  
}
\ffigbox{  \includegraphics[width=0.45\textwidth]{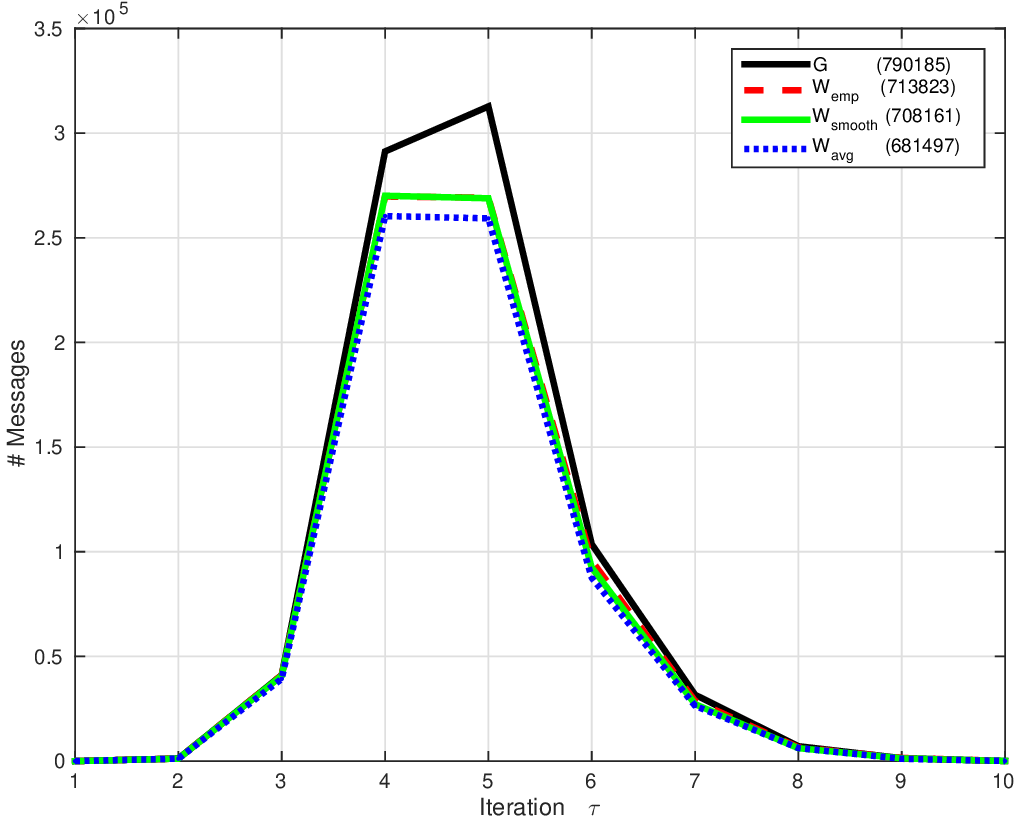} }
{%
\caption{Average number of messages per iteration (YouTube; totals in brackets) }\label{f:average_message_cnt_youtube}
  
}

\end{floatrow}
\end{figure}

Figure~\ref{f:pi_tau_theoretical_actual} shows the empirical distribution of $\pi_k^{\tau}$ (based on a sample of 500 random root nodes) for the YouTube Graph versus the theoretical (Figure~\ref{f:pi_tau_theoretical}). As can be seen for low degrees the approximation is excellent but deviates at higher degrees, especially during iteration 4. This occurs because high degree nodes in real networks cluster together in the network core (breaking the uniform assumption in the configuration model). That said, most nodes in power-law network are of low degree where the approximation is excellent and as will be seen the results are not effected adversely. 

%

\begin{figure}
 \begin{floatrow}
 
 \ffigbox{  \includegraphics[width=0.45\textwidth]{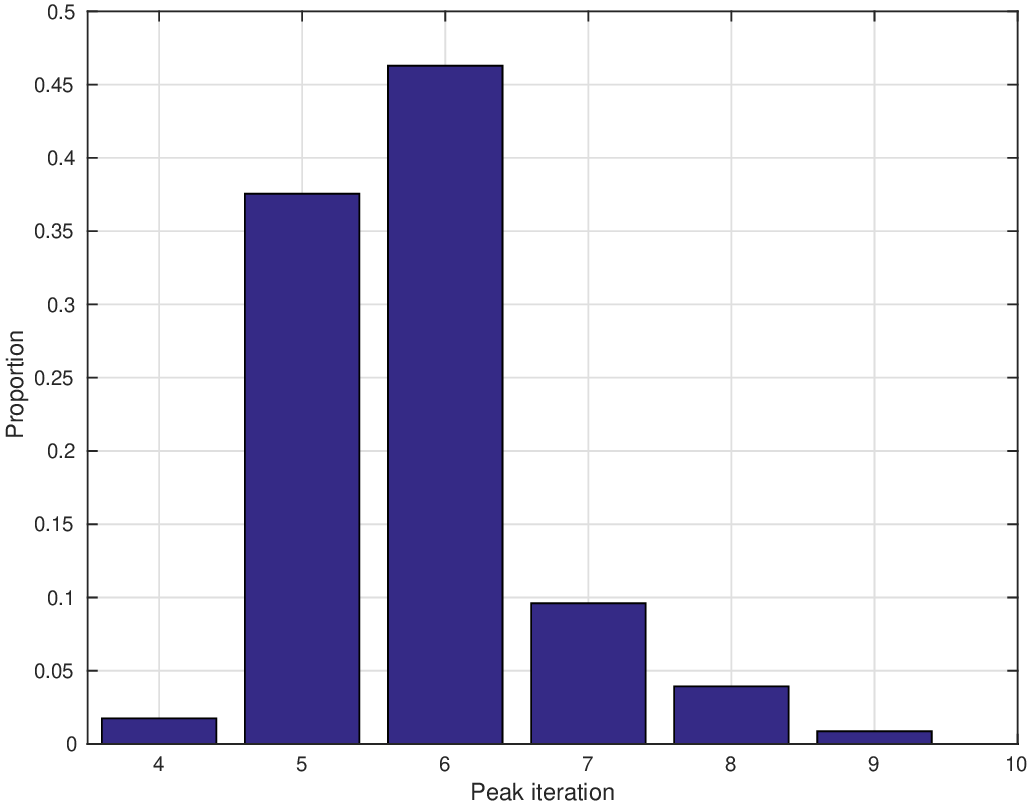} }
 {%
 \caption{Histogram of iteration at which the number of vertices in the frontier reached a peak.}\label{f:peak_histogram}
 }
   \ffigbox{  \includegraphics[width=0.45\textwidth]{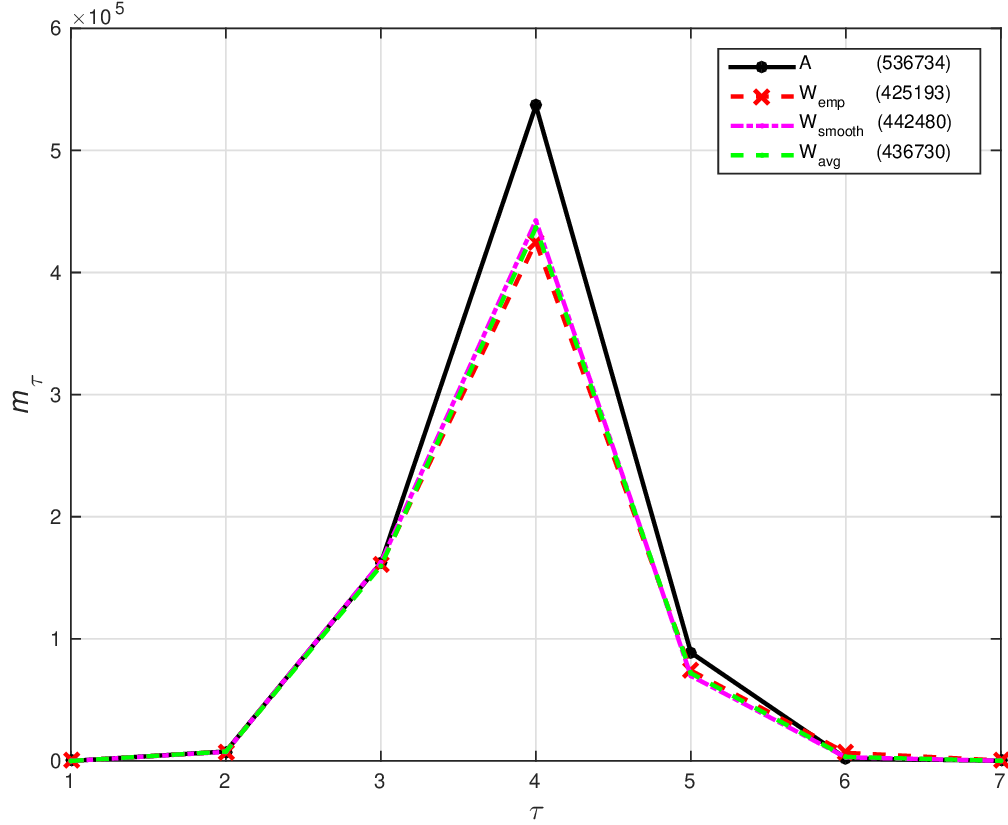} }
 {%
 \caption{Example showing number of messages per iteration for YouTube graph (root $u$=157,298) }\label{f:youtube_example_messages}
   
 }
 
 \end{floatrow}
 \end{figure}

Figure~\ref{f:average_message_cnt_youtube} shows the average number of messages per iteration using the 4 algorithms above applied to the YouTube graph. As can be seen the three weighted graph versions perform better than the unweighted graph. The average number of messages (over all iterations) transmitted using $W_{avg}$ is the lowest at 681K while those for the unweighted graph are 790K. The results differ on closer inspection however. Figure~\ref{f:peak_histogram} shows the histogram of the iteration at which the peak iteration occurred in each BFS run. For most source vertices the iteration at which the number of vertices in the frontier reaches a peak is 5 or 6.

%

\begin{figure}
 \begin{floatrow}
 
 \ffigbox{  \includegraphics[width=0.45\textwidth]{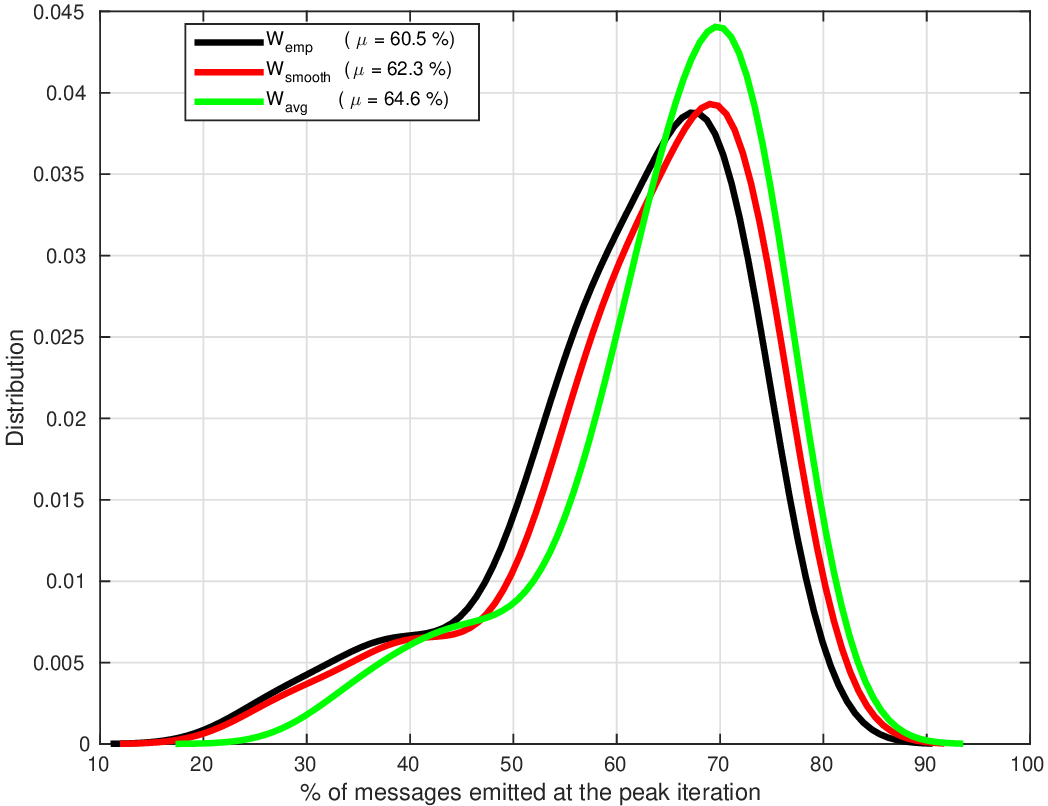} }
 {%
 \caption{The percentage of total messages emitted at the peak iteration.}\label{f:youtube_peak_percent}
 }
   \ffigbox{  \includegraphics[width=0.45\textwidth]{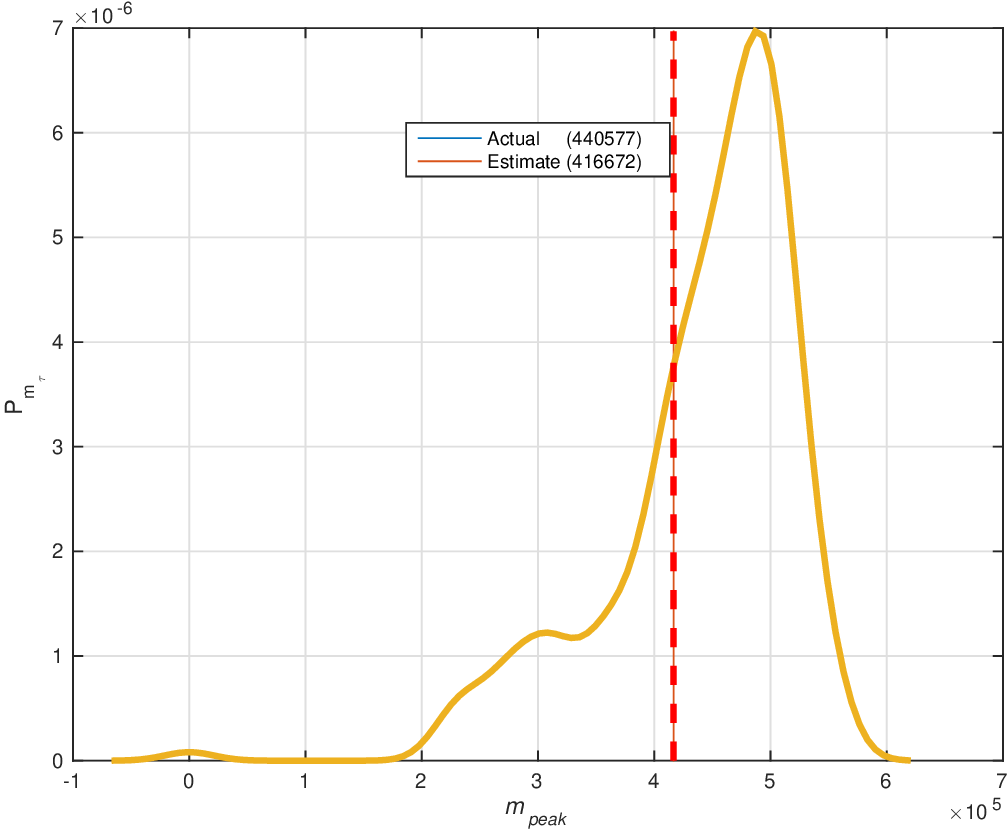} }
 {%
 \caption{Distribution of number of messages versus the expected number (YouTube graph, mean values in brackets) }\label{f:number_of_messages_partitions}
   
 }
 
 \end{floatrow}
 \end{figure}

Figure~\ref{f:youtube_example_messages} shows the distribution of messages for a particular root node and as can be seen here the peak occurs at iteration 4 and the number of messages in the peak far exceeds those in the other iterations (as is typically the case, as shown in Figure~\ref{f:youtube_peak_percent}). It is interesting to note that a weighted matrix designed to reduce the number of messages at the peak iteration should also reduce the number of messages off the peak although this is not always the case. 

%
%

Figure~\ref{f:number_of_messages_partitions} shows the distribution of the actual number of messages sent at the peak versus the estimate (Equation \ref{e:weighted_cut}). The estimate is reasonably close given it is an approximation but seems to underestimate the number of messages by about 5\%.

\begin{figure}
\begin{floatrow}
\ffigbox{  \includegraphics[width=0.45\textwidth]{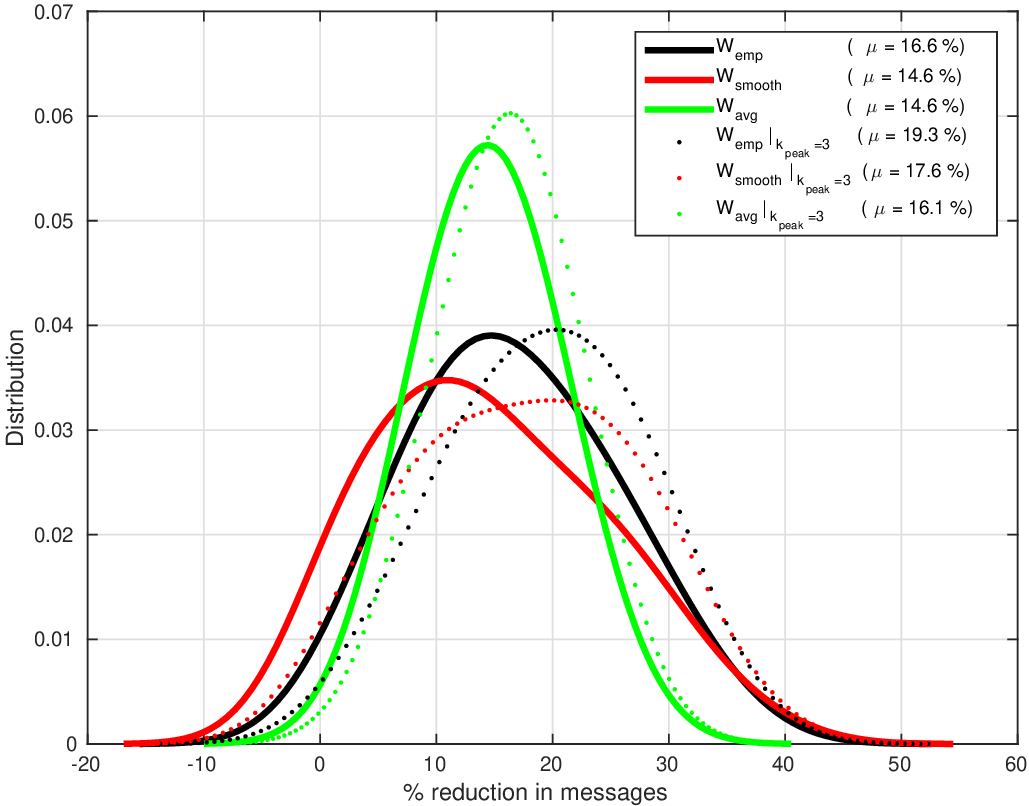} }
{%
\caption{Distribution of reduction for YouTube dataset (the distribution using those with peak $\ge 6$ is shown using the dotted line, 500 samples) }\label{f:reduction_youtube}
  
}
\ffigbox{  \includegraphics[width=0.45\textwidth]{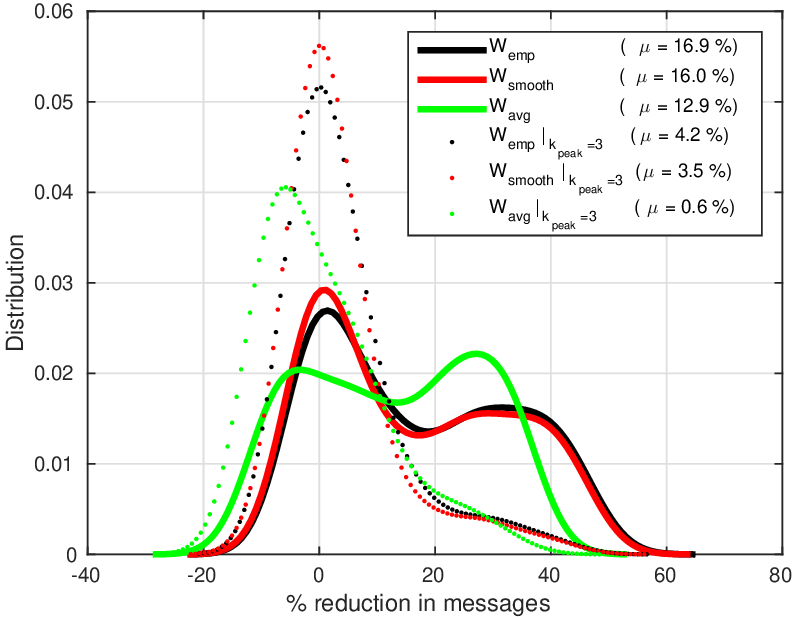} }
{%
\caption{Distribution of reduction for epinions dataset (the distribution using those with peak $\le 3$ is shown using the dotted line)}\label{f:reduction_A_e}
}

\end{floatrow}
\end{figure}

Now we turn our attention to how the algorithm \emph{performs} relative to the baseline. Figure~\ref{f:reduction_youtube} shows the \emph{percentage improvement in messages} over the the baseline algorithm. The savings are in the order of 15\% for this graph which is quite significant. In this particular case the three algorithms perform reasonably similarly but note that $W_{avg}$ leads to the lowest improvement in messages at the peak but interestingly the highest improvement in the overall number of messages (Figure~\ref{f:average_message_cnt_youtube}). 


Figure~\ref{f:reduction_A_e} shows the improvements observed with the Epinions graph. Here there is a distinct bi-modal distribution, with one distribution centred around 4\% and another centred $\sim35\%$. For this graph about half the iterations peak at $\tau = 3$ and the remainder at  $\tau = 4$. If we look at the improvement for those that peak at $\tau = 3$ alone then a clearer picture emerges. For these vertices the improvement is very small (the 4\% mode in the distribution). One possibility is that vertices which reach the peak at $\tau = 3$ lie in the core of the graph and have less hops to the periphery; thus the BFS algorithm has less time to achieve the random mixing assumed in Equation~\ref{e:wkk} (Kurant similarly notes that the starting vertex can significantly effect their estimates~\cite{Kurant}). 

%

\begin{figure}
\begin{floatrow}
\ffigbox{  \includegraphics[width=0.45\textwidth]{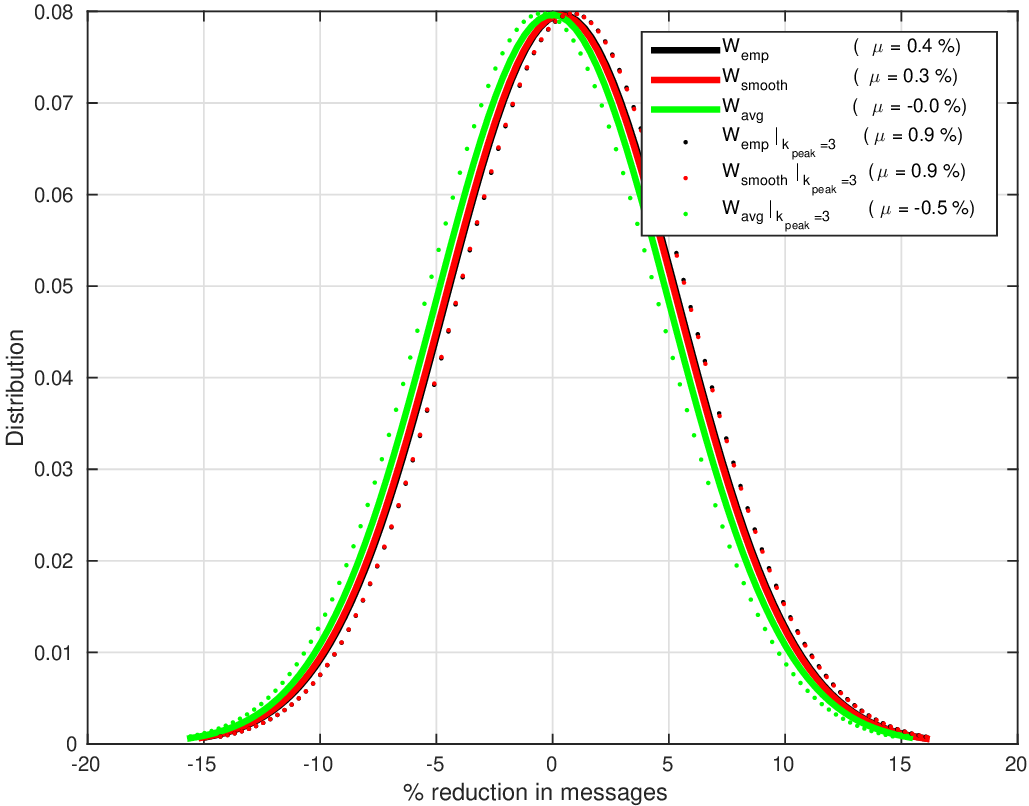} }
{%
\caption{Distribution of reduction for ER dataset (the distribution using those with peak $\ge 3$  is shown using the dotted line)}\label{f:reduction_ER}
  
}
\ffigbox{  \includegraphics[width=0.45\textwidth]{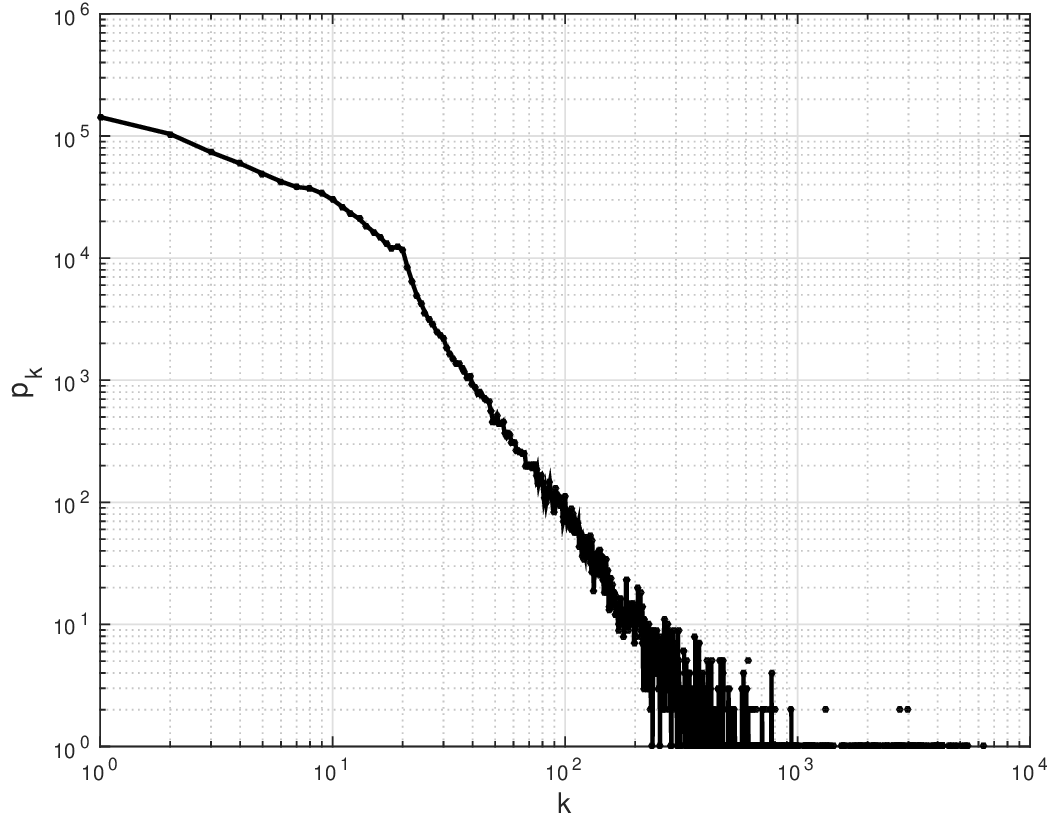} }
{%
\caption{Degree distribution for Google hyperlink graph (loglog scale)}\label{f:google_deg_dist}
  
}

\end{floatrow}
\end{figure}

Next we examine a graph with no structure, an Erdos Renyi (ER) graph, where the joint degree distribution is uniform and the degree distribution is concentrated around the mean. As there is no structure in the graph we expect the algorithm to fail and this is exactly what is seen in Figure~\ref{f:reduction_ER}.\footnote{Alternatively one could insert a concentrated degree distribution for $p_k$ in (\ref{e:f_k}) and see that $\pi_{k}^{tau} = p_k$.} The \% (dis)improvement is a distinctive Gaussian distribution centred on zero. 


Moving onto a collection of graphs, Table~\ref{t:congruence} summarizes our results. These results are quite mixed; for some graphs the reduction in messages can be very significant and in the order of $\sim15\%$ while for others it can be quite low. For the epinions and YouTube graphs the improvement is 12.80\% and 14.59\% on average which is not far from the upper bound of 16.90\% and 16.57\%. For the Catster, Wikipedia, and DBLP graph the results are reasonable and in the region of 5\% (3.8\%, 4.8\%, 6.7\%). The result for the Google Hyperlink graph is less promising and in fact shows that the algorithm degrades performance! Investigating further we found that the degree distribution for this graph is not power law. Figure~\ref{f:google_deg_dist} displays the distinctive power law tail but the distribution for low degree nodes is more uniformly distributed breaking the underlying assumption required for the algorithm to work. 

For the Epinions graph the result for the non-core vertices increases to 17.20\% but for the YouTube graph it actually decreases to 12.37\%. For the DBLP graph, there is no difference. For Wikipedia the difference is quite significant with non-core vertices reporting a reduction in messages up from  4.80\% to 13.96\%. The main conclusion here is that the position of a vertex in the graph certainly has an effect on the performance but it is unclear what the effect will actually be.

\begin{table*}[ht] 
{

\scriptsize 
    \centering
    \resizebox{\columnwidth}{!}{
    \begin{tabular}{|l|c|c|c|c|c|c|c|}
      \hline
 Name&     Type  &     $|V|$   &   $|E|$    &    $r$  & $\rho_{emp}$ \%  &  $\rho_{smooth}$ \%  &  $\rho_{avg}$ \%  \\
 \hline
YouTube Friendship & Social & 1,134,890  &  2,987,624 & -0.03 & 16.57 (12.59) & 14.61 (10.18) & 14.59 (12.37)\\
Epinions & Social & 75,879 &  508,837 & -0.04 & 16.9  (21.4) & 15.9 (20.3) & 12.8 (17.2) \\
Gowalla & Social & 196,591& 950,3279 &-0.02 &11.79 (10.38) &9.52 (8.02) & 7.2 (6.62)\\
DBLP & Coauthorship & 1,314,050 & 18,986,618 & 0.10 & 8.75 (8.76) &  8.11 (7.99) & 6.74 (6.56) \\   
Wikipedia En & Hyperlink & 1,853,493  &  39,953,145 & -0.05 &7.75 (15.15) & 6.69 (13.62) & 4.80 (13.96) \\  
Catster Friendship & Social & 149,700 &  5,449,275 & -0.16 & 4.62 (3.09) & 3.51 (2.36)&  3.86 (2.61) \\         
Google & Hyperlink & 875,713 & 5,105,039& -0.05 &-3.14 (-3.75)&-0.83 (-0.34)&-0.52 (-0.89)\\   
ER graph & Synthetic & 100,000 & 1,151,281 & 0.00 & 0.41 (0.23) & 0.34 (0.14) & -0.01 (0.16)  \\          
 
  \hline
\end{tabular}
} 
    \caption{Summary of results for a collection of graphs (http://konect.uni-koblenz.de) values for $\rho$ in brackets exclude core nodes, results averaged over 500 simulations. \label{t:congruence}}
}
\end{table*}

\section{Conclusion}
\label{sec:conclusion}
This paper has clearly demonstrated that graph structure can be leveraged to improve the efficiency of BFS; in some cases significantly by up to 20\%. In fact, any graph (not just power law graphs) with a skewed degree distribution will result in $ \pi_k^{\tau} \ne \pi_k^{\tau+1} \not\propto p_k$ and so an improved partition in theory. The computational overhead required for the algorithm to work $\{p_k, p_{k,k'}, n_\tau\}$ can be easily estimated from an initial burn in period (several BFS runs). Future work will look at extending this approach to weighted, directed graphs, we also note that as vertices and edges are added to a real-world graph its degree distribution does not change rapidly and so there is scope for application in dynamic and streaming graph analysis. The skew present in $\pi_k^{\tau}$ is such that (the standard) unweighted edge partition is not optimal for any iteration. This is why in Figure~\ref{f:average_message_cnt_youtube} we see that the total number of messages (not just at the peak) can also be radically reduced. 

%

We are currently working to implement the algorithm on a GPU and a distributed architecture. For GPU's the communications cost between processors is not uniform. In fact there exists a hierarchy with typical GPU's containing several ($\sim 15$) multi-processors, each containing several ($\sim 12$) groups, each containing ($\sim 16$) cores. Communications between cores are considered far less costly than those between multi-processors as can be used to advantage~\cite{Sungpack_HONG}.


On the surface there would not appear to be a conflict between the approach presented here and those mentioned in Section~\ref{sec:related}. Future work will investigate integration 
with these approaches for improved performance. Finally, further work is required to determine why the algorithm works better for some start vertices, if those vertices can be identified in advance, and in a computationally efficient manner. It is also possible that Equation~\ref{e:f_k} could be made conditional on known information about the root vertex. 
%



\bibliographystyle{IEEEtran}
\bibliography{bfs.bib}   

\end{document}